\documentclass{aa}
\usepackage{graphics}
\usepackage{epsfig}

\begin{document}
\title{NGC~6738: not a real open cluster}
\author{C. Boeche\inst{1}
\and   R. Barbon\inst{2}
\and   A. Henden\inst{3}
\and   U. Munari\inst{1}
\and   P. Agnolin\inst{2}
       }
\offprints{U.Munari (munari@pd.astro.it)}
\institute {
Osservatorio Astronomico di Padova, Sede di Asiago, 
I-36012 Asiago (VI), Italy
\and
Osservatorio Astrofisico di Asiago, 
Universit\`a di Padova, I-36012 Asiago (VI), Italy
\and
Universities Space Research Association/U. S. Naval Observatory
Flagstaff Station, P. O. Box 1149, Flagstaff AZ 86002-1149, USA
}
\date{Received date..............; accepted date................}

\maketitle
  
\begin{abstract}
A photometric, astrometric and spectroscopic investigation of the poorly
studied open cluster NGC~6738 has been performed in order to ascertain its
real nature. NGC~6738 is definitely not a physical stellar ensemble:
photometry does not show a defined mean sequence, proper motions and
radial velocities are randomly distributed, spectro-photometric parallaxes
range between 10 and 1600 pc, and the apparent luminosity function is identical to
that of the surrounding field. NGC~6738 therefore appears to be an apparent
concentration of a few bright stars projected on patchy background absorption.

\end{abstract}
\maketitle

\section{Introduction}

NGC~6738 ($\alpha_{2000}=19^{h}01^{m}.4$,
$\delta_{2000}=+11^{\circ}36^{\prime}$, $l=44^{\circ}.4$, $b=+3^{\circ}.1$)
shows up as a group of bright stars on a fairly crowded background located a
few degrees from the galactic equator in Aquila. It is classified as IV2p
meaning that the object is poorly populated and separated from the
surrounding field and spans a moderate range in brightness (Ruprecht
\cite{ruprecht}). No modern data exist for this object. Collinder
(\cite{collinder}) found a distance of 1190 pc and Roslund (\cite{roslund})
reported, by means of objective prism spectral classification, that the
stars down to 12 mag in the region of the cluster are dwarfs but they do not
appear to define a main sequence, thus already challenging the reality of
the object. In another study, Sahade et al. (\cite{sahade}) listed the
eclipsing variable V888 Aql as a possible cluster member. In this paper we
report on UBVRI photometry, radial velocities, spectral classifications and
Tycho-2 proper motions of stars in the field of the cluster, in order to
ascertain its true nature. As a matter of fact, looking at the Palomar Atlas
maps, the cluster region seems to show up more like a window of low
absorption in a larger region of high obscuration than as a physical
grouping of stars.
 
\section {Photometry}

\begin{table}[b]
\begin{center}
\begin{tabular}{cccc}
\hline
Date & Exposures & Field  & Seeing \\
yymmdd &        & ($\prime$) & ($\prime\prime$) \\ 
\hline
990720 & short & 11x11 &  2.0\\
990720 & medium & 11x11 &  2.0\\
990720 & long & 11x11 &  2.0\\
990721 & short & 11x11 &  1.8\\
990721 & medium & 11x11 &  1.8\\
990721 & long & 11x11 &  1.8\\
000729 & medium & 44x44 & 3.0\\
\hline
\end{tabular}
\end{center}
\caption{Journal of observations.
{\em Seeing} is the FWHM of stellar images as measured on the CCD frames.
Short exposures are a few seconds; medium exposures are around a minute;
and long exposures are several minutes in duration.}
\label{J_Obs}
\end{table}

\begin{figure*}[!t]
\vbox{
\resizebox{11.5cm}{!}{\includegraphics{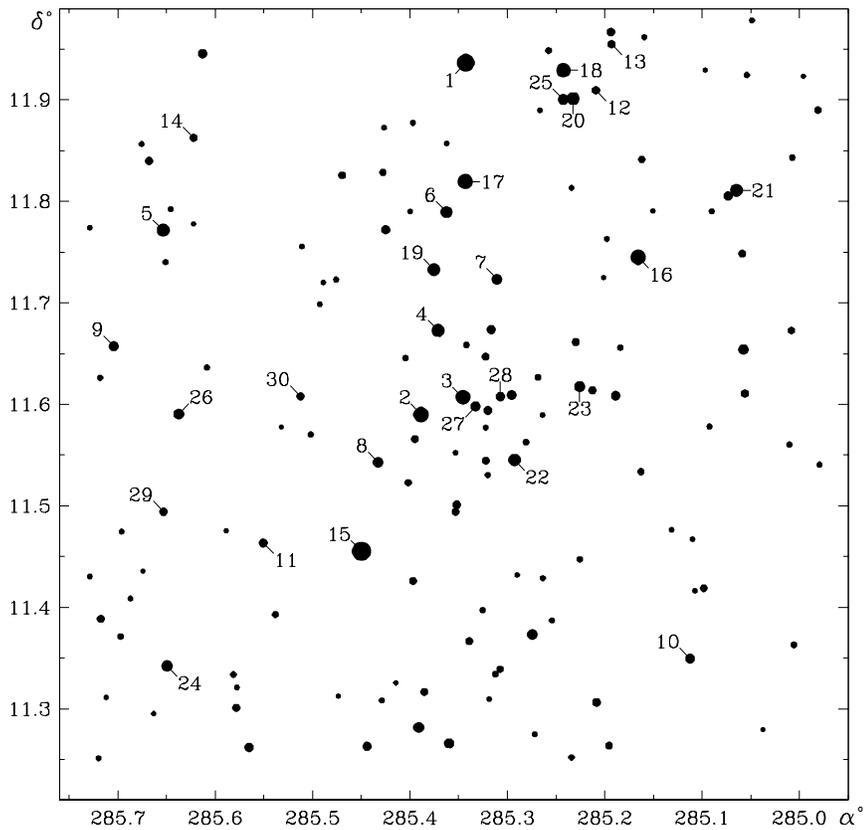}}}
\vspace{-2.6cm}
\hfill\parbox[t]{5cm}{\caption[]{Finding chart of the stars brighter than
V=13 in the field of NGC~6738. Dots scale with magnitude. Stars
spectroscopically observed are numbered.}}
\label{f_chart}
\end{figure*}

\noindent
Photometric observations were made with the 1.0-m Ritchey-Chr\'etien
telescope of the U. S. Naval Observatory, Flagstaff Station, in two
successive runs on 1999 and 2000. The journal of observations is given in
Table~\ref{J_Obs} and a finding chart for the brighter stars is presented in
Figure~\ref{f_chart}.  In the first run the surveyed area was
$11.4\times11.4$ arcmin$^{2}$ centered on the cluster position and with a
limiting magnitude of V = 20.0, whereas in the second run a larger area of
$44\times44$ arcmin$^{2}$ was imaged down to V=17.0 in order to have a
larger scale sampling of the field surrouding the cluster.

A Tektronix/SITe 1024x1024 thinned, backside--illuminated CCD was used for
the first epoch and a Tektronix/SITe 2048x2048 CCD for the second epoch,
along with Johnson UBV and Kron--Cousins RI filters. Images were processed
using IRAF, with nightly median sky flats and bias frames.  Aperture
photometry was performed with routines similar to those in DAOPHOT (Stetson
\cite{stetson}). Astrometry was performed using SLALIB (Wallace
\cite{wallace}) linear plate transformation routines in conjunction with the
USNO--A2.0 reference catalog. Errors in coordinates were typically under 0.1
arcsec in both coordinates, referred to the mean coordinate zero point of
the reference stars in each field. The telescope scale is 0.6763 arcsec/pixel.
Typical seeing was $\sim$2 arcsec. A 9 arcsec extraction aperture with
concentric sky annulus was commonly used. The reported photometry only uses
data collected under photometric conditions (transformation errors under
0.02 mag).  Cluster observations were interspersed with observations of
Landolt (\cite{landolt}, \cite{landolt1}) standard fields, selected for wide
colour and airmass range.  The mean transformation coefficients (cf. Henden
\& Kaitchuck \cite{henden}, eqns. 2.9ff) are:

\begin{figure*}[!ht]
\centering
\resizebox{16.6truecm}{!}{\includegraphics{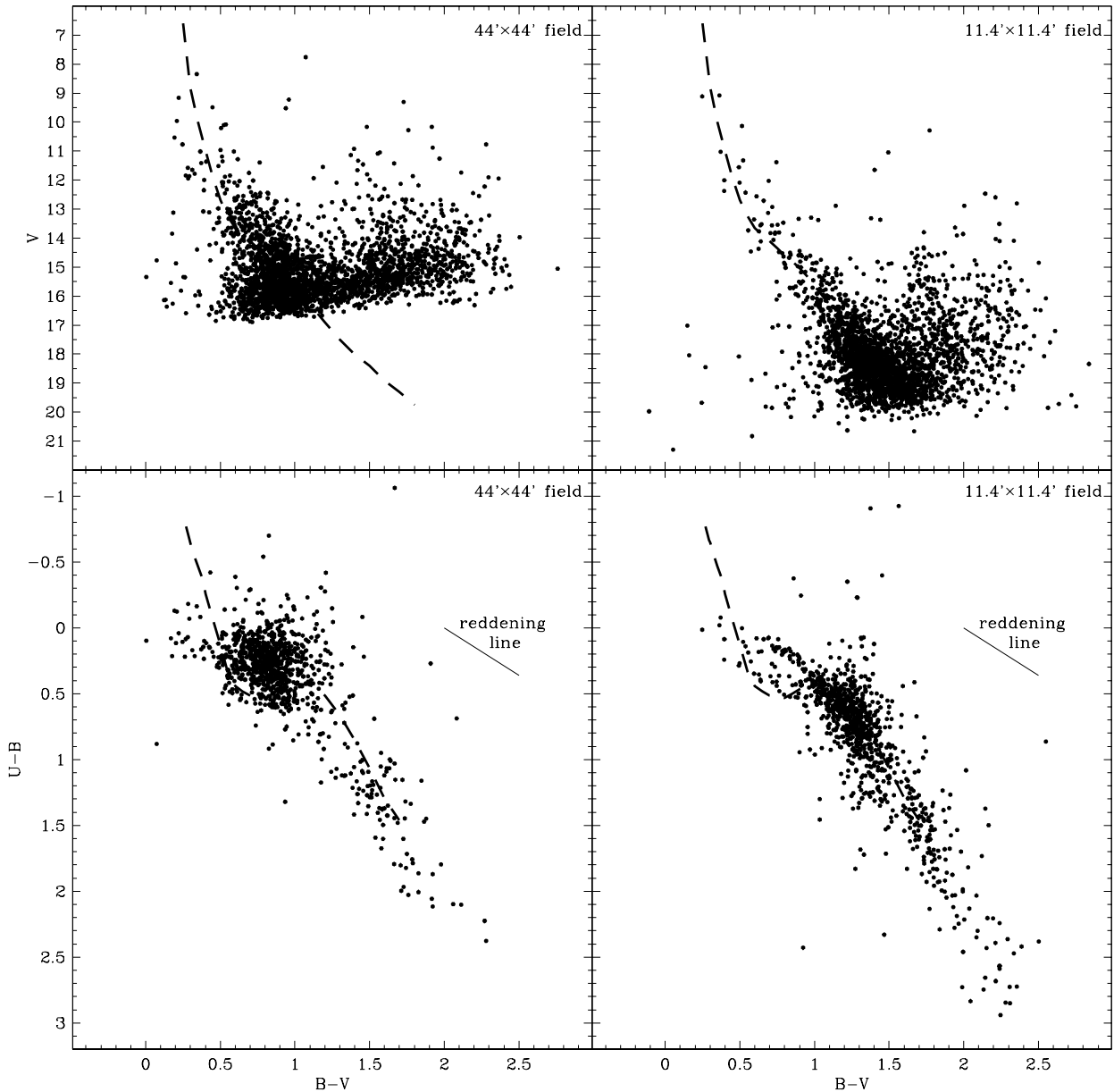}}
\caption{ V,(B-V) diagrams (top) and  (U-B),(B-V) diagrams (bottom) 
of the 44$\times$44 arcmin$^2$ field  (left) and 11.4$\times$11.4 
arcmin$^2$ field (right). The position of Blaauw ZAMS (dashed lines) for m-M=12.1
and  E(B-V)=0.6 is marked. }
\label{CMD_CCD}
\end{figure*}

\begin{eqnarray}
  V:& -0.020& \pm 0.007 \\
B-V:& ~0.949& \pm 0.007 \\
U-B:& ~1.072& \pm 0.018 \\
V-R:& ~1.017& \pm 0.005 \\
R-I:& ~0.971& \pm 0.013 
\end{eqnarray}

\noindent
Second order extinction was negligible except for  \mbox{B-V}, where a
coefficient of --0.03 was used. The resulting UBVRI magnitudes and position
of the observed stars are available in electronic form
\footnote{\tt ftp://ftp.nofs.navy.mil/pub/outgoing/aah/sequence/ /ngc6738.dat}.
The V,(B-V) and (U-B),(B-V) diagrams for the stars observed are shown
in Figure~\ref{CMD_CCD}.  Assuming a slope of 0.72 for the reddening line in
the colour-colour (U-B),(B-V) diagrams and a Blaauw (1963) ZAMS in
both colour-colour and colour-magnitude V,(B-V) diagrams (hereafter CMD), we
tentatively fit at the same time all the main sequences, obtaining a mean
reddening of E(B-V)=0.6 and a distance modulus of $(m-M)=12.1$. But the apparent
pattern of a main sequence must not deceive us. In fact, as shown by Burki
and Maeder (\cite {burki}), a main sequence is not proof of the presence
of an open cluster: field stars can trace a fictitious main sequence. In
order to discriminate field stars from cluster stars, Burki and Maeder
suggest:
\begin{description}
\item[(a)] observing the lower limiting envelope of the main
sequence in CMDs (field star fictitious sequences are steeper than those
of real clusters).
\item[(b)] observing the position of the stars in CMDs
and colour-colour diagrams (field stars have no coherent positions).
\item[(c)] observing the luminosity
function (unlike cluster stars, the field star apparent luminosity function
always reaches its
maximum at the limiting magnitude). 
\end{description}
\noindent

\begin{figure}[!t]
\resizebox{\hsize}{!}{\includegraphics{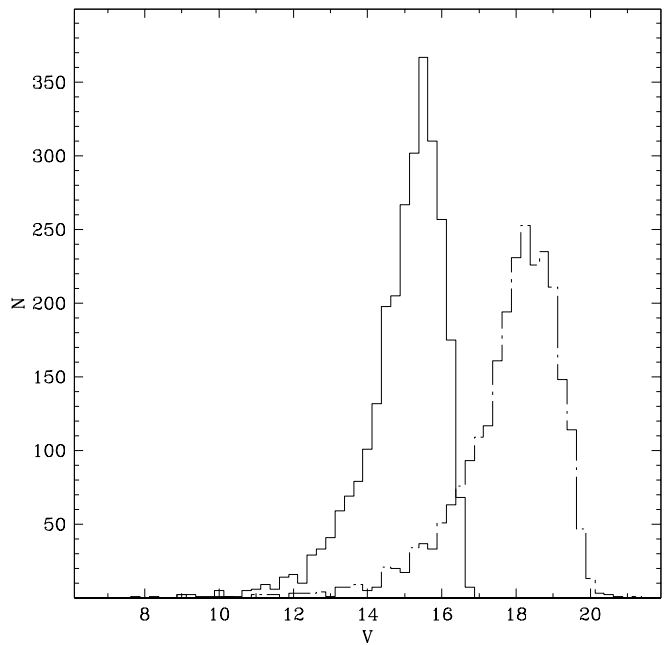}}
     \caption{Apparent luminosity function for the 44$\times$44 arcmin$^2$ 
(solid line) and for the 11.4$\times$11.4 arcmin$^2$ (dot-dashed line).}
\label{L_func}
\end{figure}

Looking at point (a), at the top of Figure~\ref{CMD_CCD} we see that the
lower limiting envelope of the main sequences differs from the Blaauw's ZAMS
at the fainter magnitudes. But if an open cluster should exist, it could be
hidden by the field stars, so this argument is not decisive.  Point (c)
is tested in Figure~\ref{L_func} where the apparent luminosity functions of
both fields decrease abruptly at their magnitude limit: this is a typical
feature of field stars. In the case of a cluster, the luminosity
function should show a secondary peak at brighter magnitudes, which is not
present in our data. Comparing the diagrams of both fields in
Figure~\ref{CMD_CCD}, the contamination due to main sequence and red giant
stars lying on the galactic disk is apparent. The structure of the CMDs
is reminiscent of a pattern composed of different stellar populations, at different
distances and different reddenings along the line of sight, as for example
reported by Ng et al. (\cite{ng}) and Bertelli et al.  (\cite{be}) in their
investigation of the galactic structure towards the galactic centre.

\begin{table*}[!tb]
\vbox{
\begin{tabular}{rlrrlllrr} 
\hline
star & TYC & V    & B-V   & \multicolumn{2}{c}{spectral type} & $E_{B-V}$ 
                  & (B-V)$_0$ & $V_{\circ}-M_{V}$ \\
\cline{5-6}
&number & &   & this paper    & literature &  &  \\
\hline
1    & HIP 93402     &  8.35  & 0.34  & A4 V         & A0     & +0.23\phantom{:}  & +0.11\phantom{5:}   &  5.91\phantom{:}  \\ 
2    & 1048 00548 1  &  9.11  & 0.25  & A1 III       & A      & +0.24\phantom{:}  & +0.01\phantom{5:}   &  8.18\phantom{:}  \\ 
3    & 1048 00812 1  &  9.08  & 0.36  & F2 IV/V      & F0     & +0.01\phantom{:}  & +0.35\phantom{5:}   &  6.04\phantom{:}  \\ 
4    & 1048 00652 1  & 10.09  & 0.54  & F7 IV/V      & F5     & +0.05             & +0.49\phantom{5:}   &  7.08\phantom{:}  \\ 
5    & 1048 01244 1  & 10.10  & 0.52  & F5 V         & A7     & +0.08\phantom{:}  & +0.44\phantom{5:}   &  6.34\phantom{:}  \\ 
6    & 1048 00663 1  & 10.53  & 0.19  & A0 III       & A0     & +0.22\phantom{:}  &--0.03\phantom{5:}   &  9.84\phantom{:}  \\ 
7    & 1047 00340 1  & 11.01  & 0.59  & G0 V         &        & +0.01\phantom{:}  & +0.58\phantom{5:}   &  6.58\phantom{:}  \\ 
8    & 1048 00726 1  & 11.02  & 0.37  & A3 III       &        & +0.29\phantom{:}  & +0.08\phantom{5:}   &  9.61\phantom{:}  \\ 
9    & 1048 00898 1  & 11.19  & 0.51  & F5 IV/V      & F5     & +0.07\phantom{:}  & +0.44\phantom{5:}   &  7.97:            \\ 
10   &               & 11.28  & 0.62  & F8 V         &        & +0.10\phantom{:}  & +0.52\phantom{5:}   &  6.98\phantom{:}  \\ 
11   &               & 11.58  & 0.28  & A2 V         &        & +0.23\phantom{:}  & +0.05\phantom{5:}   &  9.56\phantom{:}  \\ 
12   & 1047 02770 1  & 11.84  & 0.27  & A3 V         &        & +0.19\phantom{:}  & +0.08\phantom{5:}   &  9.75\phantom{:}  \\ 
13   & 1047 02104 1  & 11.86  & 0.29  & A0 III       &        & +0.32\phantom{:}  &--0.03\phantom{5:}   & 10.86\phantom{:}  \\ 
14   &               & 11.95  & 0.28  & A0 III       &        & +0.31\phantom{:}  &--0.03\phantom{5:}   & 10.97\phantom{:}  \\ 
15   & 1048 00908 1  &  7.76  & 1.07  & G6 IV        & K0     & +0.25             & +0.82\phantom{5:}   &  3.82\phantom{:}  \\ 
16   & 1047 01117 1  &  9.23  & 0.96  & G8 V         & K0     & +0.22\phantom{:}  & +0.74\phantom{5:}   &  3.05\phantom{:}  \\ 
17   & 1048 00949 1  &  9.30  & 1.73  & K7 V         & M0     & +0.40\phantom{:}  & +1.33\phantom{5:}   &--0.03\phantom{:}  \\ 
18   & 1047 01033 1  &  9.52  & 0.94  & G8 V         & K0     & +0.20\phantom{:}  & +0.74\phantom{5:}   &  3.41\phantom{:}  \\ 
19   & 1048 00192 1  & 10.16  & 1.92  & K7 V         & M0     & +0.59\phantom{:}  & +1.33\phantom{5:}   &  0.24\phantom{:}  \\ 
20   & 1047 01087 1  & 10.16  & 1.48  & K4 III       & K5     & +0.10\phantom{:}  & +1.38\phantom{5:}   &  9.85\phantom{:}  \\ 
21   & 1047 00370 1  & 10.21  & 0.50  & F3 V         & A5     & +0.12\phantom{:}  & +0.38\phantom{5:}   &  6.22\phantom{:}  \\ 
22   & 1047 00861 1  & 10.29  & 1.77  & M1 III       & M      & +0.19\phantom{:}  & +1.58\phantom{5:}   & 10.20\phantom{:}  \\ 
23   & 1047 02069 1  & 10.77  & 0.25  & A1 V         & A0     & +0.24\phantom{:}  & +0.01\phantom{5:}   &  9.03\phantom{:}  \\ 
24   & 1048 00104 1  & 10.77  & 2.28  & K5 III:      &        & +0.78:            & +1.50:\phantom{5}   &  8.55:            \\ 
25   & 1047 02012 1  & 10.92  & 1.40  & K3 V         &        & +0.44\phantom{:}  & +0.96\phantom{5:}   &  2.92\phantom{:}  \\ 
26   & 1048 00660 1  & 10.96  & 0.50  & A3 V         & A0     & +0.42\phantom{:}  & +0.08\phantom{5:}   &  8.16\phantom{:}  \\ 
27   & 1048 00360 1  & 11.05  & 1.49  & K5           &        & +0.34:            & +1.15:\phantom{5}   &  2.63:            \\ 
28   & 1047 01773 1  & 11.38  & 0.74  & G6 IV        &        & --0.06:           & +0.82\phantom{5:}   &  8.45:            \\ 
29   &               & 11.75  & 0.69  & F4 IV        &        & +0.27             & +0.42\phantom{5:}   &  8.38\phantom{:}  \\ 
30   & 1048 00582 1  & 11.89  & 0.53  & A4 V         &        & +0.42             & +0.11\phantom{5:}   &  8.87\phantom{:}  \\ 
\hline
\end{tabular}
\vspace{-8.4cm}
}
\hfill\parbox[t]{4cm}{\caption[]{Spectroscopic data. Column from left to
right: star number, Tycho (or Hipparcos) identification,
V magnitude, B-V colour, our spectral types and those from
literature (if avaiable), E(B-V), (B-V)$_0$ and distance moduli. Colour
excesses and distance moduli have been computed using Lang's (1992) tables.
Colon indicates uncertainties in the intrinsic colours and/or absolute
magnitudes. The ``negative'' reddening shown by star \#28 could trace
an undetected binary.}}
\label{Sp.data}
\end{table*}

\section{Spectroscopy}

\begin{figure}[!t]
\resizebox{\hsize}{!}{\includegraphics{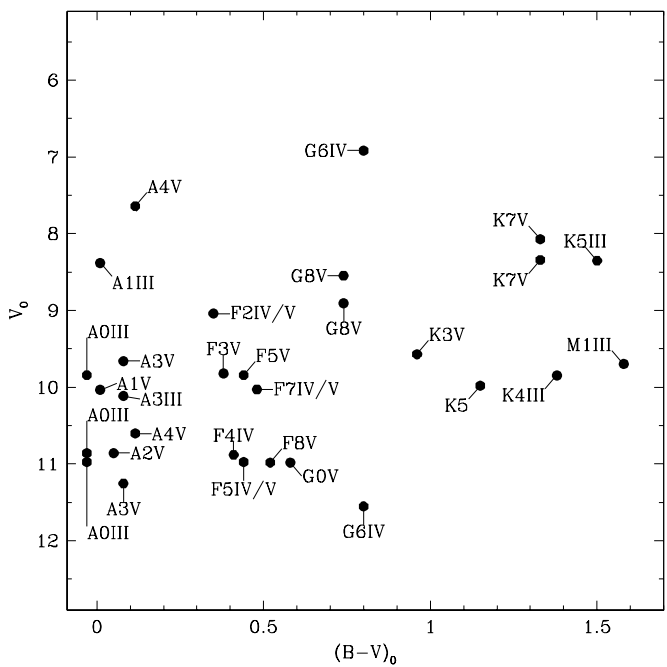}}
     \caption{Intrinsic colour-magnitude diagram of observed stars, 
each of them labeled with its spectral classification.}   
\label{CMD}
\end{figure}

For classification purposes, spectroscopic observations of 30 stars in a
$44\times44$ arcmin$^{2}$ area centered on the cluster have been obtained
with the B\&C spectrograph at the 122cm telescope of the Asiago Observatory.
A 600~gr/mm grating has been used providing a dispersion of 2 \AA/pix
(74~\AA/mm) in the 3900-4900~\AA\ interval and the observations were reduced
using IRAF.  Table~\ref{Sp.data} shows our spectral classifications
(obtained against an internal atlas of re-observed MK standards from the
list of Yamashita et al. 1977) compared to the scanty data found in the
literature together with colour excesses and corrected distance moduli
obtained from our photometry.  The stars observed appear to be distributed
between 10 to 1600 pc with no clear clustering.  This is confirmed by the
intrinsic CMD shown in Figure~\ref{CMD} where, disregarding the obvious late
type dwarfs belonging to the field, no main sequence is present. We arrive
at the same conclusion observing where the classified stars lie in the
photometric diagrams of Figure~\ref{CMD_CCD}. From point (b) in the previous
section, we must infer that all of our program stars are field stars.  This
is also supported by the reddenings of the observed stars which span about 1
mag, suggesting once more that they do not lie at the same distance.

\section {Proper motions}

\begin{figure}[!t]
\resizebox{\hsize}{!}{\includegraphics{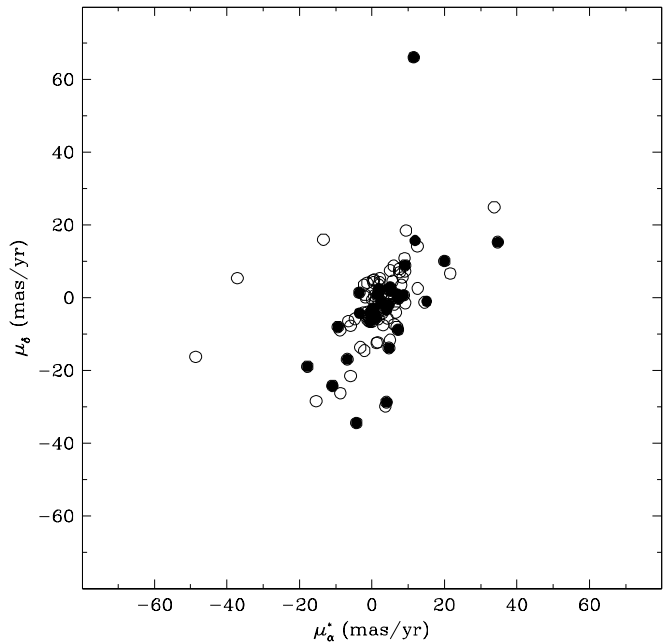}}
      \caption[]{Vector point diagram (based on Tycho-2 data) of the stars
located inside a 1$^\circ \times$1$^\circ$ area centered on NGC~6738. Error
bars do not exceed the point size. Solid ones mark the stars
spectroscopically observed.}
\label{VPD}
\end{figure}

Using proper motions from Tycho-2, a vector point diagram (VPD) of the stars
located inside a box of 1$^{\circ}\times$1$^\circ$ on the center of the
cluster is given in Figure~\ref{VPD}. This diagram shows no concentration of
cluster stars distinct from the distribution of field stars (cf Lattanzi et al. 1991).
From this characteristic, we can infer the absence along the line of sight
to NGC~6738 of a localized group of stars sharing the same projected space motion. 

\section {Radial velocities}

\begin{table}[!b]
\begin{center}
\begin{tabular}{crc} 
\hline
star    & $RV_{\odot}$~~ &  JD \\
	& km/sec~       &  (+2451000)\\
\hline
   &                      &           \\
 1 &      --{\bf 7}.6$\pm0.9$    &   798.287 \\ \cline{2-3}
 2 &       5{\bf 8}.8$\pm1.8$   &   773.375 \\
   &       3{\bf 1}.0$\pm1.6$  &   798.379 \\
   &       3{\bf 8}.2$\pm2.2$  &   799.373 \\ \cline{2-3}
 3 &     --3{\bf 6}.9$\pm1.3$    &   773.333 \\
   &     --3{\bf 4}.7$\pm1.0$  &   773.440 \\ \cline{2-3}
 4 &     --5{\bf 6}.3$\pm0.1$   &   773.354 \\ \cline{2-3}
 5 &      --{\bf 5}.3$\pm0.7$   &   799.458 \\ \cline{2-3}
 6 &     --3{\bf 6}.1$\pm0.5$   &   798.359 \\ \cline{2-3}
 8 &     --3{\bf 4}.1$\pm1.8$   &   773.391 \\
   &     --4{\bf 1}.8$\pm1.0$  &   798.399 \\
   &     --2{\bf 0}.8$\pm1.0$  &   799 430 \\ \cline{2-3}
11 &     --3{\bf 5}.7$\pm0.4$   &   798.420 \\ \cline{2-3}
22 &     --5{\bf 8}.9$\pm0.4$    &   773.419 \\
   &                      &           \\
\hline
\end{tabular}
\end{center}
\caption{Radial velocities from Echelle spectra. The listed errors are
errors of the mean.}
\label{RVtable}
\end{table}

\noindent

Some of the brightest stars in the field of the cluster have been observed
with the Echelle spectrograph mounted at the 182 cm telescope operated by Osservatorio Astronomico di Padova 
at Mount Ekar (Asiago, Italy).
The set up used provided a 20\,000 resolving power over the 4550 - 8750~\AA\
interval. Data reduction and analysis has been performed with IRAF. 
Table~\ref{RVtable} lists the heliocentric radial velocities of program
stars together with their errors as estimated from the comparison of the
results obtained from the various echelle orders. The errors cluster around
1~km~sec$^{-1}$. Stars \#2 and \#8 show large radial velocity variations that
suggest a binary nature.  As for the preceding section, the large scatter of
radial velocities does not support the presence of a real cluster. Three
stars group around RV =$-$35 km~sec$^{-1}$ and 9.5$\leq$m--M$\leq$10.0 mag (stars
\#6, \#8 and \#11). Star \#3 shows the same radial velocity but it lies at a
closer distance.

\section{Stellar counts}

\begin{figure}[!t] 
\psfig{figure=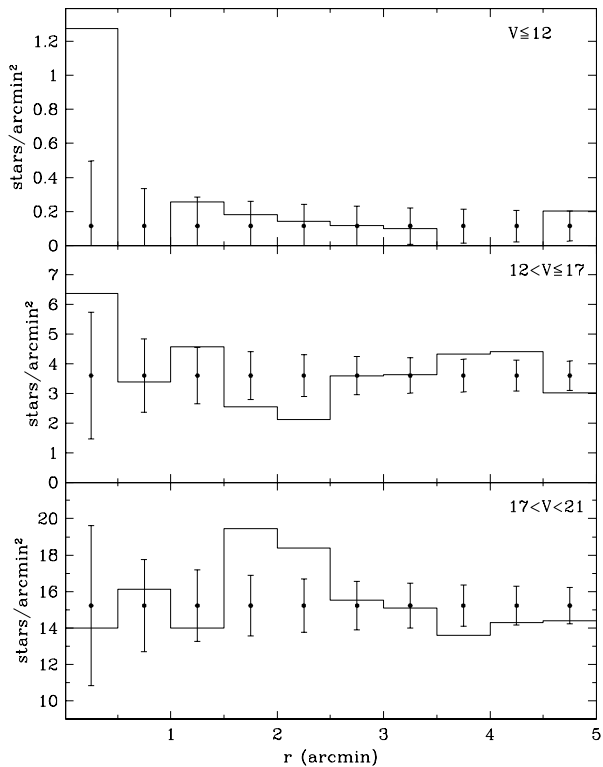,width=8.0cm,clip=}
      \caption[]{Histogram of the surface density of the stars in the 
11.4$\times$11.4 arcmin$^2$ field of NGC~6738 in three different magnitude ranges.
Bins show the stellar density in 0.5 arcmin$^2$ wide annuli. Dots indicate the
expected mean with 1$\sigma$ errorbars provided by Poissonian statistics.}
        \label{annuli}
\end{figure}

In order to test the physical reality of an open cluster, it is common to
analyse its radial density distribution. In real clusters the density should
decrease outwardly, eventually merging with the surrounding field. Following
Odenkirchen and Soubiran (\cite{soubiran}), we have measured the stellar
density in 0.5 arcmin$^2$ wide annuli centered at the nominal position of
NGC~6738 for three different magnitude ranges. The result
(Figure~\ref{annuli}) shows a stellar density distribution in agreement with
Poissonian statistics (represented by dots with error bars) except for
$V\leq$12 mag stars of the inner circle (histogram on top of
Figure~\ref{annuli}) and which we have spectroscopically observed and
already found not to be physically related.

\begin{figure}[!t]
\resizebox{\hsize}{!}{\includegraphics{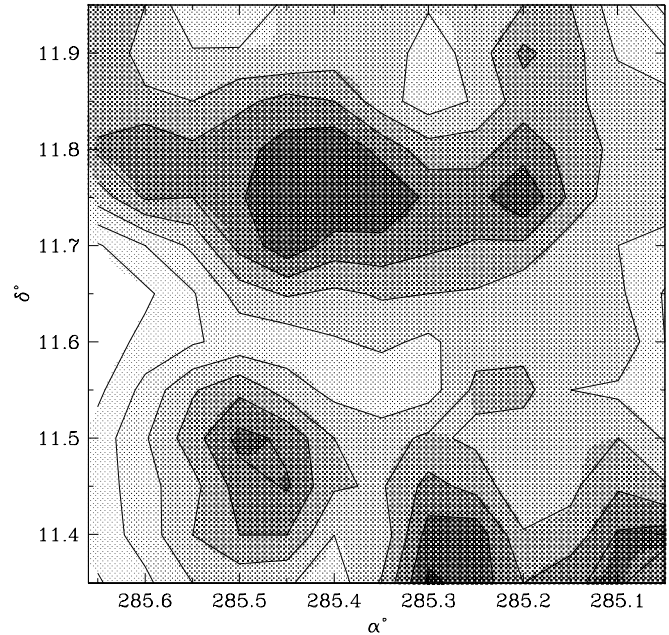}}
      \caption[]{Surface stellar density from 44$\times$44 arcmin$^2$ field. 
Grey levels in steps of 0.25~stars/arcmin$^2$. It covers a range from 
0.5~stars/arcmin$^2$ (lightest) to 2.3
stars/arcmin$^2$ (darkest).
              }
        \label{density}
   \end{figure}

\noindent
Figure~\ref{density} shows the integrated stellar density distribution in
the field of NGC~6738. The highest density is off-centered with respect to
the cluster position, and the overall pattern is very patchy. The pattern
is the mirror image of the dust emission at 100~$\mu$m measured by IRAS
(cf. http://irsa.ipac.caltech.edu/applications/ISSA/): the emission
concentrates towards the areas of lowest stellar counts. This fact is indicative
of an extremely non homogeneous distribution of interstellar material over the
field of view. Coupled with a chance grouping of a few bright foreground
stars, this concentration could erroneously lead to a cluster detection.

\section{Conclusions}

Our purpose was to perform an investigation as complete as possible to
verify the existence of the open cluster NGC~6738. Tycho-2 proper motions have
been combined with our new deep and wide-field UBVRI photometry, radial velocities
and spectral classification. We have not found evidence supporting the
existence of a real cluster; the colour-magnitude and colour-colour diagrams
do not show a reliable cluster main sequence; the spectro-photometric
parallaxes of the 30 brightest stars
show no concentration in distance.; the apparent luminosity function is 
one of field stars; proper motions and radial velocities do not support a
common space motion of the program stars; the stellar density distribution in the
field of NGC~6738 does not have a peak in the supposed location of the
cluster, nor a negative gradient moving away from it, as 
expected if a real cluster were present, but instead it reveals the patchy structure
of the interstellar absorption confirmed by the IRAS data.  Finally, the
concentration of stars brighter than $V$=12 mag toward the center of the field, from
which one may infer the presence of a cluster and certainly drove the
earlier investigators to pick up this object, has been shown 
to be a chance grouping. Concluding, it appears safe
to conclude that NGC~6738 is not a real open cluster.

\begin{acknowledgements}
We thank G.Bertelli for useful discussions {\bf and B.Skiff for comments}.
\end{acknowledgements}

\end{document}